\documentclass[reprint,amsmath,amssymb,aps,pra,showkeys,superscriptaddress]{revtex4-2}

\usepackage[utf8]{inputenc}
\usepackage{graphicx}
\usepackage{dcolumn}
\usepackage{bm}
\usepackage{siunitx}
\usepackage[german,english]{babel}
\usepackage{amsmath}
\usepackage{url}
\usepackage[version=4]{mhchem}
\usepackage{xcolor}
\usepackage{romannum}

\preprint{APS/123-QED}	

\begin{document}

\title{A simple strategy to measure the contact resistance between metals and doped organic films}
\author{Anton Kirch}
\author{Axel Fischer}
\author{Robert Werberger}
\author{Shayan Miri Aabi Soflaa}
\affiliation{
	Dresden Integrated Center for Applied Physics and Photonic Materials (IAPP) and Institute of Applied Physics, Technische Universität Dresden, Germany\\
	Nöthnitzer Straße 61, 01187 Dresden, Germany 
}	
\author{Karolina Maleckaite}
\affiliation{
	Dresden Integrated Center for Applied Physics and Photonic Materials (IAPP) and Institute of Applied Physics, Technische Universität Dresden, Germany\\
	Nöthnitzer Straße 61, 01187 Dresden, Germany 
}	
\affiliation{Center of Physical Sciences and Technology, Sauletekio av. 3, LT-10257 Vilnius, Lithuania}
\author{Paulius Imbrasas}
\author{Johannes Benduhn}
\author{Sebastian Reineke}
\email{sebastian.reineke@tu-dresden.de}
\affiliation{
	Dresden Integrated Center for Applied Physics and Photonic Materials (IAPP) and Institute of Applied Physics, Technische Universität Dresden, Germany\\
	Nöthnitzer Straße 61, 01187 Dresden, Germany 
}

\date{\today}

\begin{abstract}
Charge injection from electrodes into doped organic films is a widespread technology used in the majority of state-of-the-art organic semiconductor devices. Although such interfaces are commonly considered to form Ohmic contacts via strong band bending, an experiment that directly measures the contact resistance has not yet been demonstrated. 
In this study, we use a simple metal/doped organic semiconductor/metal stack and study its voltage-dependent resistance. A transport layer thickness variation proves that the presented experiment gains direct access to the contact resistance of the device.
We can quantify that for an operating current density of $\SI{10}{mA/cm^2}$ the investigated material system exhibits a voltage drop over the metal/organic interface of about $\SI{200}{mV}$, which can be reduced by more than one order of magnitude when employing an additional injection layer.
The presented experiment proposes a simple strategy to measure the contact resistance between any metal and doped organic film without applying numerical tools or elaborate techniques. Furthermore, the simplistic device architecture allows for very high, homogeneous, and tunable electric fields within the organic layer, which enables a clear investigation of the Poole-Frenkel effect. 
\end{abstract}

\maketitle

\section{Introduction}

Doped organic layers enable efficient charge carrier injection and extraction in various emerging semiconductor devices and have a huge impact on their overall performance. They enable reduced voltage losses at the contact interfaces of photovoltaic (PV) devices \cite{guoSulfamicAcidCatalyzedLead2016,trukhanovEffectDopingPerformance2011,xiaoContactEngineeringElectrode2017,dingReducingEnergyLosses2020}, account for the low operating voltage and reasonable charge balance of light-emitting diodes (LEDs) using organic injection layers \cite{yahyaEffectsArgonPlasma2021,khanElucidatingImpactCharge2019,weiHighlyStableEfficient2022}, and enhance the switching speed of organic transistors \cite{waldripContactResistanceOrganic2020,dollingerVerticalOrganicThinFilm2019}. Thus, physical understanding and easy measurement routines are crucial to evolving this technology.

Any charge injection or extraction between electrodes and organic semiconductors experiences a resistance at the contact interface. Its mechanism is commonly treated in the picture of a metal-semiconductor bilayer rooting in classical semiconductor physics and is standard content in pertinent textbooks \cite{szePhysicsSemiconductorDevices2021,kohlerElectronicProcessesOrganic2015,forrestOrganicElectronicsFoundations2020}. Even though amorphous organic semiconductors do not form energy band structures in the sense of their crystalline inorganic counterparts, the model of valence and conduction band is often applied to the distribution of molecular orbitals in amorphous organic films for the sake of simplicity and delivers a readily elaborate physical understanding. In this picture, charge carriers face a potential barrier caused by the difference between the metal work function and the semiconductor's energy levels. Tuning this injection barrier is commonly achieved by energy level alignment, i.e. by introducing injection layers \cite{kumataniPracticalChargeInjection2013,kumakiReducingContactResistance2008,kotadiyaUniversalStrategyOhmic2018}, or energy band bending \cite{kohlerElectronicProcessesOrganic2015, kahnElectronicStructureElectrical2003,aurouxEvidenceEffectsIon2021}. 

The latter option is mediated via blending the organic semiconductors with dopant molecules \cite{walzerHighlyEfficientOrganic2007,tietzeFermiLevelShift2012,tietzeElementaryStepsElectrical2018} and provides an enhanced tunneling probability for charge carriers, which is even increased by an image charge potential \cite{limketkaiChargeInjectionCathodedoped2005,hosseiniChargeInjectionDoped2005}. Depending on the intended majority charge carrier, an electron or hole transport layer requires n-type or p-type doping, respectively. Organic molecules, such as \ce{F4-TCNQ} and \ce{F6-TCNNQ}, or halide materials like \ce{FeCl3} are commonly used as p-type dopants. Alkali metals like \ce{Li} or \ce{Cs}, on the other hand, are very popular n-type dopants \cite{bruttingPhysicsOrganicSemiconductors2012}.

Under standard operating conditions, the interface between metal and doped organic semiconductor is commonly treated as Ohmic and only poses a significant contact resistance below certain bias voltages \cite{liaptsisCrosslinkableTAPCBasedHoleTransport2013,zufleDeterminationChargeTransport2017,siemundNumericalSimulationOrganic2016}. Naturally, the question arises what the term ``Ohmic'' refers to. And indeed, its definition is not intuitive and sometimes even confused with a connection to Ohm's law. In classical semiconductor physics, the term ``Ohmic'' describes a contact that poses negligible resistance relative to other resistances in the device regardless of the applied bias polarity and may exhibit non-linear behavior \cite{szePhysicsSemiconductorDevices2021,kohlerElectronicProcessesOrganic2015}. Is it actually correct that an Ohmic contact can be assumed at common organic LED or PV operating conditions? How much voltage drop does the contact resistance cause? The answer is not straightforward and depends on the application intention of the respective device. For high-current applications, such as future electrically driven organic lasers, these voltage losses will be severe \cite{meisterOpticallyPumpedLasing2018}. Also, considering the resistance contribution of contacts may develop the understanding of OLED operating voltages deviating from their theoretical limits \cite{meerheimHighlyEfficientOrganic2006,meerheimInfluenceChargeBalance2008}, can pave tracks enhancing transistor performance \cite{borchertCriticalOutlookPursuit2022}, and provide further insight into voltage losses in organic or perovskite PV devices \cite{sandbergRelatingChargeTransport2016}.

While there have been extensive research efforts to understand metal/doped organic semiconductor contacts in detail \cite{sandbergRelatingChargeTransport2016,altazinAnalyticalModelingContact2011,oehzeltOrganicSemiconductorDensity2014}, this study presents a direct experimental method to measure the contact resistance of such material combinations. It is a particular challenge to isolate the pure contact behavior, as it is commonly overlaid with the charge transport characteristics of the investigated device. To overcome this problem, we reduce the device complexity as much as possible: A doped organic semiconductor with variable thickness is sandwiched between two equally thick silver electrodes and its voltage-dependent device characteristics are studied. As a result, the resistance contributions are reduced to the two contacts and a transport layer, whose influence is controlled via a thickness variation. 
At the same time, we drastically decrease the active area and use a 4-wire crossbar setup to measure high current densities unaffected by parasitic series resistance or self-heating, which would otherwise be detrimental to the experimental outcome.
By thoroughly studying the presented model system, we can prove that the manufactured devices follow classical semiconductor physics to a fair extend and that the introduced experimental strategy gains direct access to the contact characteristics of a metal-organic semiconductor interface.

\section{Results}
\subsection{Device architecture}
\begin{figure}[!t]
	\includegraphics[width=1.0\linewidth]{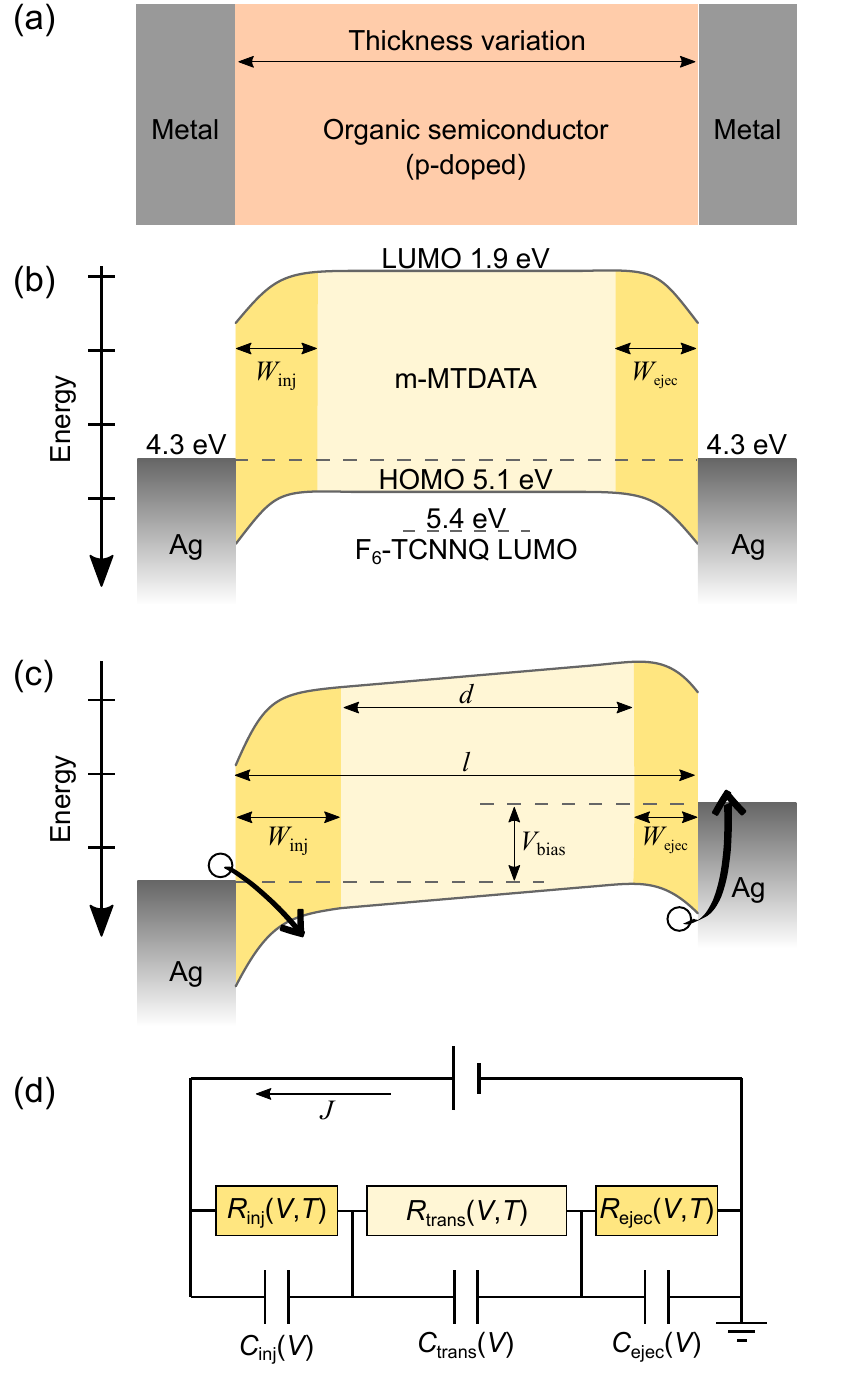}
	\caption{(a) Device design: A simple metal/doped organic semiconductor/metal stack with a variation of semiconductor thickness. (b) Energy level diagram at short circuit: The p-doping induces band bending and depletion regions at the interfaces. Energy levels taken from Refs. \cite{aonumaMaterialDesignHole2007,michaelsonWorkFunctionElements1977, koechSynthesisApplication8Hexafluorotetracyanonaphthoquinodimethane2010}. (c) Energy level diagram under applied bias. (d) Equivalent circuit model: The physical characteristics of the presented stack can be described by a simple series of contact and transport equivalent circuits, each represented by a parallel RC element.}
	\label{figure_stack_circuit}
\end{figure}

Figures \ref{figure_stack_circuit}(a) and \ref{figure_stack_circuit}(b) present the device architecture with a p-doped organic semiconductor  (\ce{m-MTDATA}:\ce{F6-TCNNQ}, $\SI{4}{wt\percent}$) evaporated between two silver electrodes. The doping ratio was chosen high enough to yield a low charge carrier injection/extraction barrier, but also low enough to sustain a medium doping efficiency \cite{olthofPhotoelectronSpectroscopyStudy2009,tietzeFermiLevelShift2012}. The semiconductor thickness is varied from $\SI{50}{nm}$ to $\SI{400}{nm}$ to study the impact of the charge transport resistance. This strategy is similar to the conception of a transmission line method (TLM) experiment often performed with transistors \cite{reevesObtainingSpecificContact1982,rheeMetalSemiconductorContact2008,horowitzInterfaceModificationTuning2011}. The crucial differences are, first, a drastically reduced device complexity sporting no dependence on geometry or charge carrier accumulation. Second, the semiconductor thickness in our experiments is orders of magnitudes below a horizontal organic transistor channel length. The transport resistance becomes, as shall be presented, almost insignificant for the thin devices used in our experiments. This enables a direct and voltage-dependent contact resistance measurement without extrapolation to zero semiconductor thickness as performed in TLM. 

The minimal transport layer is chosen to be $\SI{50}{nm}$, as films below this thickness become increasingly prone to short-circuit failures. Furthermore, any spatial overlap of the two depletion regions must be avoided to keep the device physics clear. With typical values for our material system like the relative permittivity of a typical organic semiconductor $\epsilon_{\mathrm{r}}\approx 3$ \cite{bruttingPhysicsOrganicSemiconductors2012}, the intrinsic potential barrier at the interface $V_{\mathrm{BI}}=\SI{0.5}{eV}$, and the ionized acceptor concentration $N_{\mathrm{A}}=\SI{1e18}{cm^{-3}}$, the depletion width $W$ of one contact can be estimated at $V=0$ to range around $\SI{4}{nm}$.

\begin{align}
	W=\sqrt{\frac{2\epsilon_0\epsilon_{\mathrm{r}}(V_{\mathrm{BI}}-V)}{eN_{\mathrm{A}}}}\approx \SI{4}{nm}\label{equ_depl_width}
\end{align}

The depletion width $W$ additionally depends on the applied bias, as schematically indicated in Fig. \ref{figure_stack_circuit}(c). Still, a minimum transport layer thickness of $\SI{50}{nm}$ safely rules out depletion region interference. 

The utilization of a single doped organic film guarantees a very homogeneous current density and electric field distributions within the device, which is not compromised by additional intrinsic layers or organic-organic interfaces in the film \cite{zhengExploitingLateralCurrent2019,fischerSelfheatingEffectsOrganic2012}. This is proven by current density over voltage ($JV$) measurements for different sample areas, which were all found to coincide closely, cf. SI Sect. \Romannum{4}. For the experiments presented throughout this study, we used the smallest realized cross-section area of $\SI{0.09}{mm^2}$ to resolve the lowest possible current densities and to prevent self heating-induced inhomogeneities scaling with active area dimension \cite{kirchElectrothermalTristabilityCauses2021}. 

\subsection{Resistance measurements}

\begin{figure*}[t!]
	\includegraphics{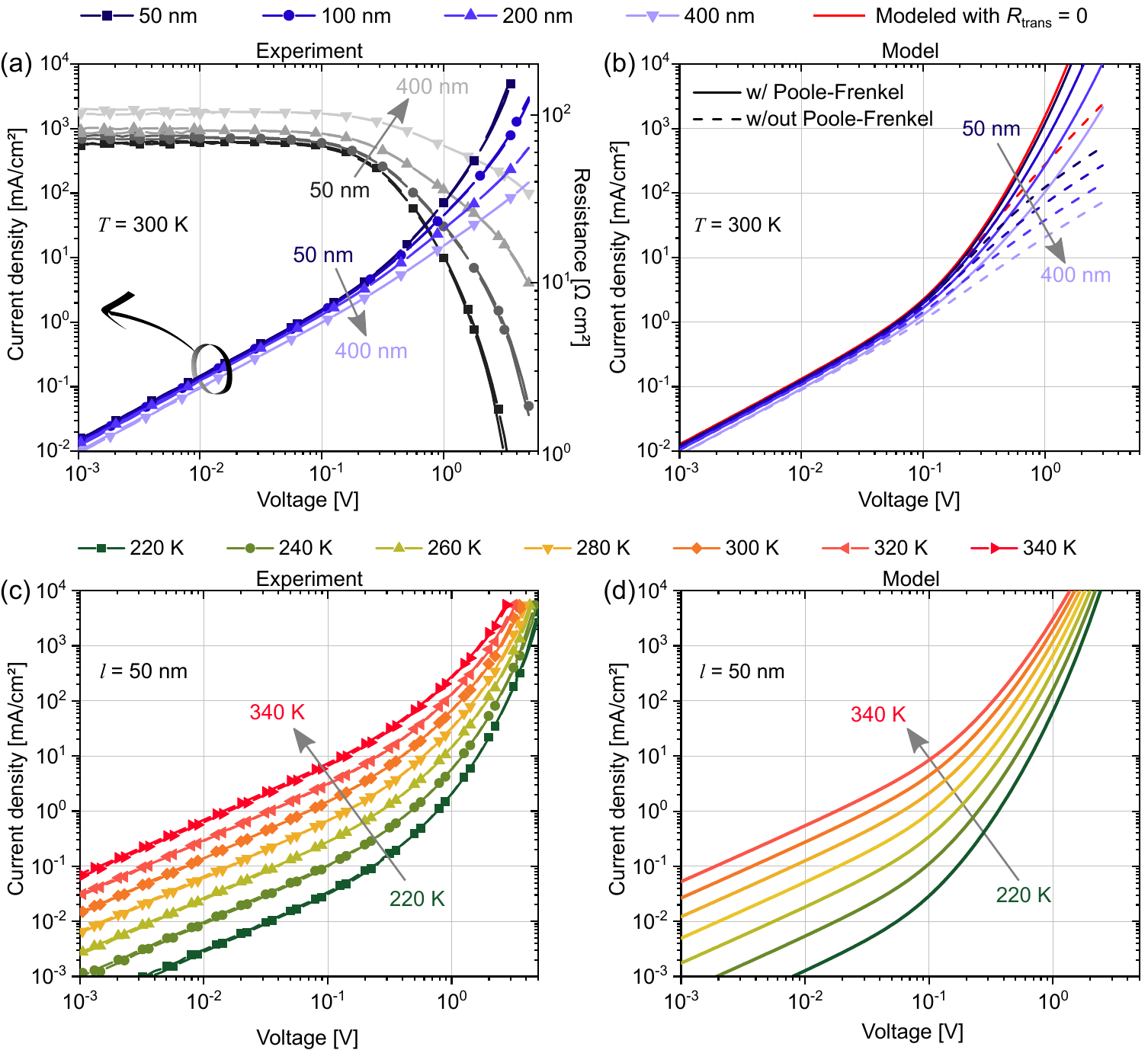}
	\caption{Panels (a) and (b) show direct current $JV$ characteristics scanned back and forth at $T=\SI{300}{K}$ for a thickness variation of the organic semiconductor film. Panel (a) also presents its $RV$ representation. Panels (c) and (d) show direct current $JV$ characteristics for the $l=\SI{50}{nm}$ device at various temperatures. Panels (a) and (c) present experimental results, panels (b) and (d) data from the equivalent circuit modeling.}
	\label{figure_IV}
\end{figure*}

A four-wire $JV$ scan is run back and forth in a Peltier element-based cryostat under vacuum (below $\SI{1}{mbar}$) to ensure stable thermal conditions. Details about the four-wire setup are introduced in Ref. \cite{kirchExperimentalProofJoule2020}. The technique enables more accurate measurements excluding external series resistances that jeopardize a precise capacitance analysis. This is essential for observing the intended characteristics. 

Figure \ref{figure_IV}(a) presents the measurement results for all four devices at a constant temperature $T=\SI{300}{K}$. The blue-shaded lines show the $JV$ characteristics and the grayscale curves the same data set in resistance representation $R=V/J$. Only a very slight hysteresis, which is scaling with increasing current density, is detected. It can be identified more clearly in Fig. S2. Electrothermal feedback, which causes a deviation between backward and forward $JV$ measurement directions, can therefore be rendered unimportant for the overall data analysis \cite{kirchExperimentalProofJoule2020}.

For $V<\SI{300}{mV}$, the $JV$ curves of the four different transport layer thicknesses differ only slightly. This indicates that they are governed by the contact resistance between metal and doped organic semiconductor at low voltages. If the transport resistance posed a major contribution, the transport layer variation by a factor of 8 would cause a respective line separation. This is not the case. While the $\SI{400}{nm}$ device shows a slight current density and resistance deviation, the three thinnest devices can merely be kept apart. 

For $V>\SI{300}{mV}$, the resistance of all devices plummets. Also, the lines start separating drastically according to their transport layer thickness, i.e. the contact resistance drops below the transport resistance which starts to govern the device characteristics. 

Figure \ref{figure_stack_circuit}(d) presents an equivalent circuit that can be used to better understand this behavior. Any of the three introduced device layers, i.e. \textit{injection}, \textit{transport}, and \textit{ejection} layer, needs to comprise a resistive component. The circuit model also emphasizes that our devices sport two contact resistances, one at each metal/organic film interface. The experiment, therefore, measures the sum of a forward and reverse contact resistance contribution.

Ejection and injection interfaces each need to be represented by a variable resistor $R_\mathrm{inj}$ and $R_\mathrm{ejec}$, as all devices are limited by their contact resistances at low bias voltages. 
Instead of an accurate but complex interface analysis, only the governing physical concepts are included. This is sufficient to understand the device physics, as we see that our model successfully represents all experimental trends. Accurate quantitative tracing of the experimental results is not our intention here. The charge carrier injection and ejection currents can be understood as Schottky contacts and are described by a tunneling term \cite{wangCarrierMobilityOrganic2010} and an ideal diode equation \cite{szePhysicsSemiconductorDevices2021}, which is reduced by the effective, i.e. image force-reduced, potential barrier at the interface $V_\mathrm{BI}$.

\begin{align}
	j_\mathrm{Diode}=j_\mathrm{0,S}\exp\left(\frac{-eV_\mathrm{BI}}{n_\mathrm{i}k_\mathrm{B}T}\right)\left[\exp\left(\frac{eV}{n_\mathrm{i}k_\mathrm{B}T}\right)-1\right]\label{equ_diode}\\
	j_\mathrm{Tun}=\sigma_\mathrm{0,R}V\exp\left(\frac{-eV_\mathrm{BI}}{n_\mathrm{i}k_\mathrm{B}T}\right)\left[\exp\left(c\frac{\sqrt{|V|}}{k_\mathrm{B}T}\right)-1\right]\label{equ_tun}
\end{align}

The parameter $k_\mathrm{B}$ depicts the Boltzmann constant, $T$ the device temperature, $n_\mathrm{i}$ the diode ideality factor, $e$ the elementary charge, $c$ an arbitrary tunneling scaling factor, $V$ the voltage drop across the respective resistor element, $j_\mathrm{0,S}$ the scale current density, and $\sigma_\mathrm{0,R}$ the scale conductivity. Please note that $\sigma_\mathrm{0,R}\cdot V$ again yields a current density, which is the current density of the tunneling term. The global model parameters are given in Tab. \ref{tab_paramters}.

The symmetric architecture yields a symmetric device behavior, as shown in the SI Fig. S2. So, both contacts need to be described by the same formula. We choose a simple superposition, which only differs in the respective signs for injection and ejection current densities.
\begin{align}
	j_\mathrm{inj}=j_\mathrm{Tun}(V_\mathrm{inj})-j_\mathrm{Diode}(-V_\mathrm{inj})\\
	j_\mathrm{ejec}=-j_\mathrm{Tun}(-V_\mathrm{ejec})+j_\mathrm{Diode}(V_\mathrm{ejec})\label{equ_jejec}
\end{align}

Both contact elements are susceptible to the voltage drop across the respective interface and decrease exponentially, as found by the experimental data. An increasing bias causes a more pronounced band bending at the injection interface, which increases the tunneling probability. Also, the Schottky barrier at the ejection interface is reduced \cite{pahnerPentaceneSchottkyDiodes2013}. Both processes make the total contact resistance drop with increasing bias voltage.

The transport resistance must also be represented by a variable resistor element $R_\mathrm{trans}$, since the devices depend on the thickness of the organic film at elevated bias voltages. Due to the high doping concentration throughout the entire organic semiconductor film, space charge-limited effects cannot play a pivotal part and the transport resistance is assumed to be purely Ohmic. Please be referred to the SI Sect. \Romannum{2} for the detailed reasoning. The current density passing $R_\mathrm{trans}$ reads
\begin{align} 
	j_\mathrm{trans}=e\,N_\mathrm{A}\mu\frac{V_\mathrm{trans}}{d}\label{equ_trans},
\end{align}
with $	d = l-W_\mathrm{inj}-W_\mathrm{ej}$ being the thickness of the non-depleted transport layer and $l$ the total thickness of the doped organic film, cf. Fig. \ref{figure_stack_circuit} (c), $V_\mathrm{trans}$ the voltage drop across the transport layer, and $N_\mathrm{A}$ temperature-activated ionized dopant concentration, equal to the effective charge carrier density, cf. SI Sect. \Romannum{1} and the parameters in Tab. \ref{tab_paramters}.
\begin{align}
	N_\mathrm{A}=N_\mathrm{A,0}\exp\left(\frac{-E_\mathrm{A,a}}{k_\mathrm{B}T}\right)\label{equ_doping_conc}
\end{align}

As temperature- and field-dependent experiments are carried out and alter the device characteristics, the mobility term $\mu=\mu(T,V)$ must contain a temperature and an electric field dependency to successfully model the experimental results. $R_\mathrm{transport}$ has to drop with increasing bias, since the $JV$ curves bend upwards when separated according to their transport layer thickness. This cannot be explained by linear resistance characteristics. The behavior at high electric fields, here exceeding $\SI{100}{\kilo V/cm}$, requires a field-induced mobility enhancement, commonly known as Poole-Frenkel (P-F) effect \cite{wangCarrierMobilityOrganic2010,kohlerElectronicProcessesOrganic2015}.
\begin{align}
	\mu = \mu_\mathrm{0}\exp\left(\frac{-E_\mathrm{\mu,a}}{k_\mathrm{B}T}\right)\exp\left(\beta\sqrt{\frac{|V_\mathrm{trans}|}{d}}\right)\label{equ_mobility}\\
	\beta = \frac{e}{k_\mathrm{B}T}\sqrt{\frac{e}{\pi\epsilon_\mathrm{r}\epsilon_0}}
\end{align}

The parameter $\beta$ is defined in Ref. \cite{murgatroydTheorySpacechargelimitedCurrent1970}, $k_\mathrm{B}$ is the Boltzmann constant, $T$ the device temperature, $\epsilon_{\mathrm{r}}$ the relative permittivity of the organic film, and $E_\mathrm{\mu,a}$ the mobility activation energy. For both dependencies, temperature and electric field, we used simplistic approaches. We are well aware that both dependencies have to be modeled with more complexity if a good agreement between experimental and model results is expected. Our model fails at high electric fields, as can be seen by the increasing deviations between Figs. \ref{figure_IV}(a) and (b). This accounts for an oversimplification of the P-F term in the equivalent circuit. From the simulated results in Fig. \ref{figure_IV}(b), comprising both a scenario with and without P-F term, one can estimate the true P-F contribution to be in between the two realized modeling scenarios, i.e. a reduction of $\beta$ would yield better agreements. Pursuing this, however, is not the intention of our investigation. It is yet worth to point out that our presented experimental approach allows studying the P-F effect in organic semiconductors with utmost precision. While stacked films in complete semiconductor devices induce a varying electric field in the vertical direction, the geometry realized here ensures a very homogeneous field strength throughout the transport layer.   

The thinner the transport layer, the smaller its resistance contribution. A thinner device follows the pure contact resistance characteristics up to higher current densities than a thick device. Hence, the isolated contact resistance can be estimated best from the $l=\SI{50}{nm}$ sample. Figure \ref{figure_IV}(b) presents the model scenario with vanishing transport influence, $R_\mathrm{trans}=0$, indicated in red. It is, therefore, safe to deduce that the contact resistance characteristics must be nonlinear and stays reasonably close to the $l=\SI{50}{nm}$ device up to about $\SI{100}{mA/cm^2}$. This range is sufficient to quantify the contact resistance for most organic electronics applications.

Both contact resistance and transport characteristics are susceptible to temperature, which is presented in Fig. \ref{figure_IV}(c) and (d). That is, the mobility in eq. \eqref{equ_mobility} is enhanced by an Arrhenius term, as are the tunneling and diode characteristics in eqs. \eqref{equ_diode} and \eqref{equ_tun}. The experimental trend of temperature-induced conductivity is replicated successfully by the model.

\begin{figure*}[t]
	\includegraphics{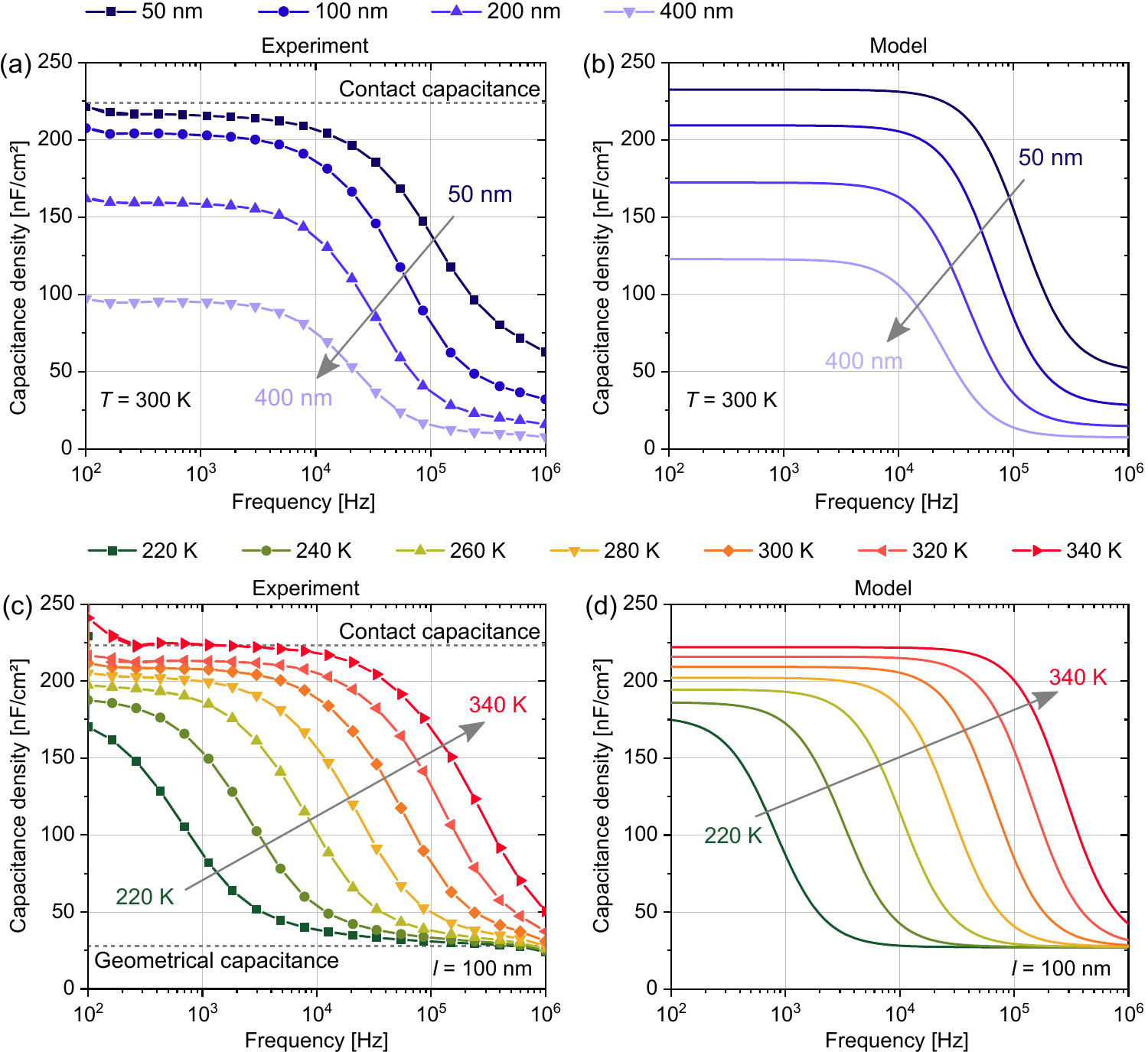}
	\caption{Impedance characteristics at $V=0$. Panel (a) and (b) show data at $T=\SI{300}{K}$ for a thickness variation of the organic semiconductor. Panels (c) and (d) show data for the $l=\SI{100}{nm}$ device at various temperatures. Panels (a) and (c) are experimental results, (b) and (d) from the equivalent circuit modeling.}
	\label{figure_Cf}
\end{figure*}

The simulation of the three resistance contributions displayed in Fig. \ref{figure_stack_circuit}(d) is evaluated with LTspice and allows to quantify each of them separately. Figure S9 presents the modeled contributions and makes it plain that the injection and ejection resistances govern the device behavior at low voltages. For $V>\SI{100}{mV}$ all three resistances start to drop dramatically. The two contact resistance contributions $R_\mathrm{inj}$ and $R_\mathrm{ejec}$, however, drop faster than the transport resistance, which takes over the device characteristics at increasing bias. Figure S9 also implies that injection and ejection follow roughly the same trend at low voltages, owing to their mutual exponential nature. Thus, an estimation to identify either the injection (important for OLEDs) or ejection resistance (important for PV) from the measurements would be $R_\mathrm{inj}\approx R_\mathrm{ejec}$ at low voltages.

\subsection{Impedance measurements}
To gather further evidence that the pure contact behavior is disclosed in the presented devices, impedance measurements are performed. Figures \ref{figure_Cf}(a) and (c) present the experimental capacitance density over frequency characteristics ($Cf$) at zero bias voltage $V=0$ with varying thickness and temperature, respectively. Please find the details on capacitance density evaluation from a four-wire measurement in the SI Sect. \Romannum{5}. Again, it is very important to employ a four-wire measuring technique to keep parasitic series resistance out of the evaluation.

\begin{figure*}[t]
	\includegraphics{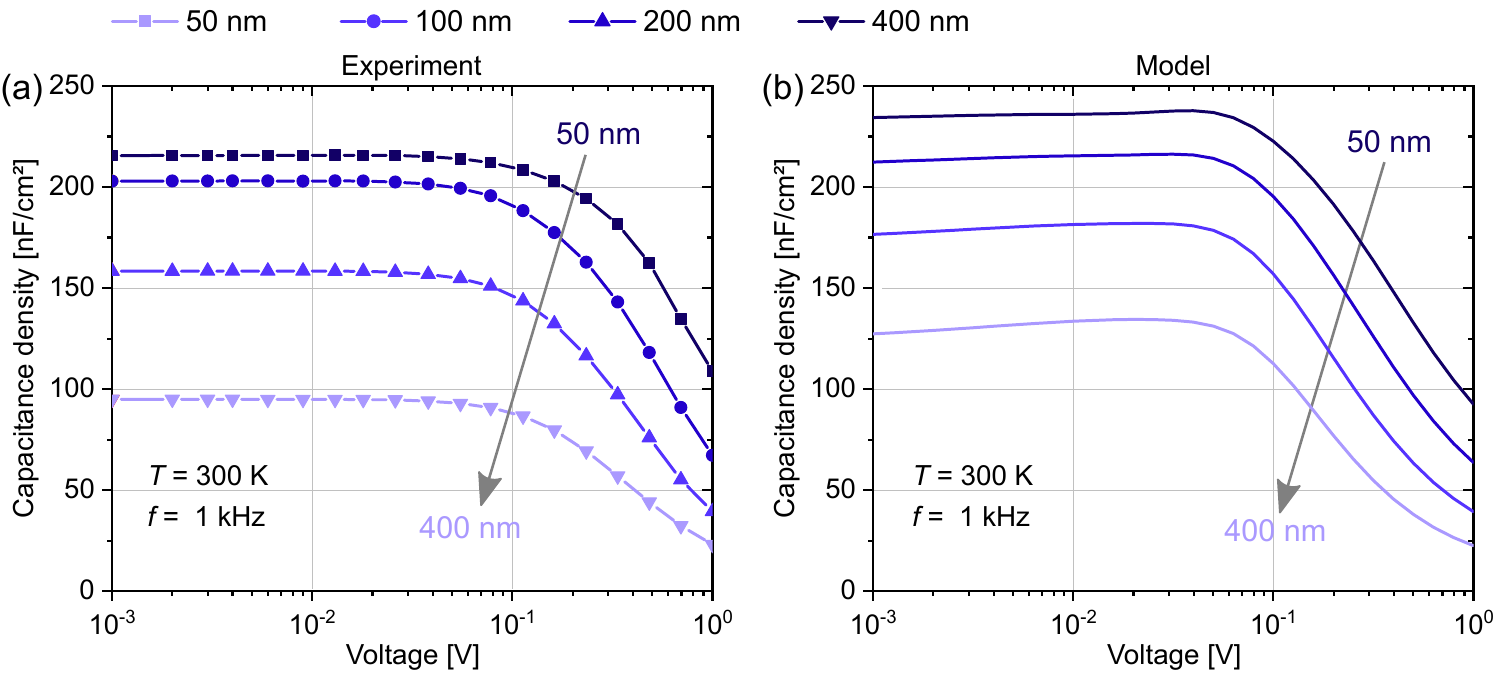}
	\caption{Impedance characteristics at $T=\SI{300}{K}$ and $f=\SI{1}{\kilo Hz}$ for a thickness variation of the transport channel. Panel (a) presents experimental data and panel (b) results from the equivalent circuit presented in Fig. \ref{figure_stack_circuit} (d).}
	\label{figure_CV-thickness}
\end{figure*}

The experimental results present two distinct capacitance density plateaus. This behavior can be understood if treating any of the three distinct layers within the device not only as having a resistive representation but as an RC element, cf. Fig. \ref{figure_stack_circuit}(d). At $V=0$, the depletion regions of injection and ejection have the same width $W$ and, treated like plate capacitors in the equivalent circuit, also the same capacitance density $C_\mathrm{inj}=C_\mathrm{ejec}$. 

\begin{align}
	C_\mathrm{inj}=\frac{\epsilon_0\epsilon_{\mathrm{r}}}{W_\mathrm{inj}}\label{equ_Cinj}\\
	C_\mathrm{ejec}=\frac{\epsilon_0\epsilon_{\mathrm{r}}}{W_\mathrm{ejec}}\label{equ_Cejec}
\end{align}

At low frequencies, the reactance of the transport capacitance $C_\mathrm{trans}$ is high and the charge carrier flow is mediated rather via $R_\mathrm{trans}$. As a result, the device capacitance is governed by the injection and ejection capacitances. At room temperature, with no bias voltage, negligible transport resistance, and low frequencies, the device capacitance density can be estimated as
\begin{align}
	C(f\ll \SI{1}{kHz}) & \approx \left(\frac{1}{C_\mathrm{inj}}+\frac{1}{C_\mathrm{ejec}}\right)^{-1}\nonumber\\
	& \stackrel{V=0}{=}\,\frac{C_\mathrm{contact}}{2},\label{equ_contact_low_f}
\end{align}
with $C_\mathrm{inj}=C_\mathrm{ejec}=C_\mathrm{contact}$ due to device symmetry at $V=0$. From Fig. \ref{figure_Cf}(a), however, it is apparent that eq. \eqref{equ_contact_low_f} holds merely for low transport resistances, i.e. very thin transport layers. The thicker the transport layer, the higher its resistance. As a result, even at low frequencies the transport capacitance mediates a significant share of the device current and reduces the total capacitance density. Thus, the capacitance at low frequencies is decreasing with increasing transport layer thickness and the pure contact capacitance can only be measured for transport layers striving toward zero thickness, i.e. the $l=\SI{50}{nm}$ device.

At high frequencies, the transport capacitance becomes dominant as its reactance drops below the resistance of the transport layer. Since the depletion region is smaller than the transport layer thickness $d$ of all devices, cf. eq. \eqref{equ_depl_width}, the contact capacitance is expected to be higher than the transport capacitance. As a result, the total capacitance density falls to a second plateau at high frequencies. This plateau is governed by the device geometry (organic film thickness) and is called the \textit{geometrical capacitance density} $C_\mathrm{geo}$. 

\begin{align}
	C(f\gg \SI{1}{kHz}) & =C_\mathrm{geo}\nonumber\\
	& = \left(\frac{1}{C_\mathrm{inj}}+\frac{1}{C_\mathrm{trans}}+\frac{1}{C_\mathrm{ejec}}\right)^{-1}
\end{align}


The transport capacitance density $C_\mathrm{trans}$ is a function of the transport layer thickness $d$ and is, as a simplistic approach, also represented by a plate capacitor in the equivalent circuit given in Fig. \ref{figure_stack_circuit}(d).

\begin{align}
	C_\mathrm{trans}=\frac{\epsilon_0\epsilon_{\mathrm{r}}}{d}\label{equ_Ctrans}
\end{align}

To support the just-presented understanding, the impedance spectroscopy was also carried out with a fixed organic film thickness ($l=\SI{100}{nm}$) for different temperatures. Figure \ref{figure_Cf}(c) presents the transition interval between the contact and the geometrical capacitance density plateau as being temperature-dependent. This can be mainly understood by considering $R_\mathrm{trans}=R_\mathrm{trans}(V,T)$ as a function of temperature according to eqs. \eqref{equ_trans} and \eqref{equ_mobility}. At low temperatures, $R_\mathrm{trans}$ is high and the current is mediated earlier via the transport capacitance, whose reactance does not scale directly with temperature. For increasing temperatures, $R_\mathrm{trans}$ decreases and the cut-off frequency $f_\mathrm{G}$ for the transport RC element increases.
\begin{align}
	f_\mathrm{G,trans}(V,T)=\frac{1}{2\pi\cdot R_\mathrm{trans}(V,T)\cdot C_\mathrm{trans}(V)}
\end{align}
Figures \ref{figure_Cf}(b) and (d) present the results from the equivalent circuit model performed with LTspice. The qualitative trends of the experiments are reproduced and even quantitatively the results are similar. The transition characteristics in the experiment, however, are sloping more gently than in the model. The equivalent circuit follows the Schottky assumption of three distinct layers having an abrupt change from depleted to neutral semiconductor characteristics. This is a simplification that allows modeling the two different, frequency-dependent operating regimes. In the real device, however, the transition between depleted and neutral regions is smooth and hence the impedance characteristics are smeared out with respect to the model.

\subsection{Depletion zone variation}
The impedance spectroscopy experiments discussed above were performed at zero bias voltage $V=0$. Here, the contact capacitances at either side of the device must be equal, according to the device's symmetry. The depletion width $W$, however, depends on the voltage drop over the respective layer, cf. eq. \eqref{equ_depl_width} and Fig. \ref{figure_stack_circuit}(c). The idea of the subsequent measurement is to investigate the influence of voltage on the contact capacitances. 

To access the isolated contact capacitance characteristics as well as possible, the experiment is performed at $f=\SI{1}{\kilo Hz}$ and room temperature, as deduced from the last section. 
Figure \ref{figure_CV-thickness}(a) presents the capacitance density of all four devices over the applied bias voltage. Using the global set of model parameters displayed in Tab. \ref{tab_paramters}, a short-circuit depletion width of about $W\approx\SI{4}{nm}$ at each interface can be expected. According to eq. \eqref{equ_depl_width} and as indicated in Fig. \ref{figure_stack_circuit}(c), $W_\mathrm{inj}$ increases and $W_\mathrm{ejec}$ decreases under forward bias. When considering the contact-limited capacitance at $\SI{1}{\kilo Hz}$, where the transport capacitance density $C_\mathrm{trans}$ is assumed to play no pivotal role, the capacitance density trend can be understood to follow
\begin{align}
	C(f=\SI{1}{kHz},V)=\left(\frac{\epsilon_{\mathrm{r}}\epsilon_0}{W_\mathrm{inj}(V_\mathrm{inj})+W_\mathrm{ejec}(V_\mathrm{ejec})}\right),\label{equ_contact_cap}
\end{align}

with both $W_\mathrm{inj}$ and $W_\mathrm{ejec}$ following eq. \eqref{equ_depl_width}. For low applied bias, both voltage dependencies apparently cancel each other, as around $V=0$ their behavior can be approximated as being linear. Beyond an applied bias of about $\SI{100}{mV}$, the linear approximation fails, and the denominator of eq. \eqref{equ_contact_cap} starts to grow. The total capacitance density of the devices drops. The same overall trend can be reproduced by the equivalent circuit simulation, presented in Fig. \ref{figure_CV-thickness}(b).

\subsection{Demonstration scenario for injection enhancement quantification}
Throughout the sections, an experimental setting was introduced that evidently provides direct access to the contact resistance of a given interface between a metal and a doped organic film.
As an outline towards potential application scenarios, this section presents a short study on the effect of additional interface layers that reduce the contact resistance.

\begin{figure}[h]
	\includegraphics{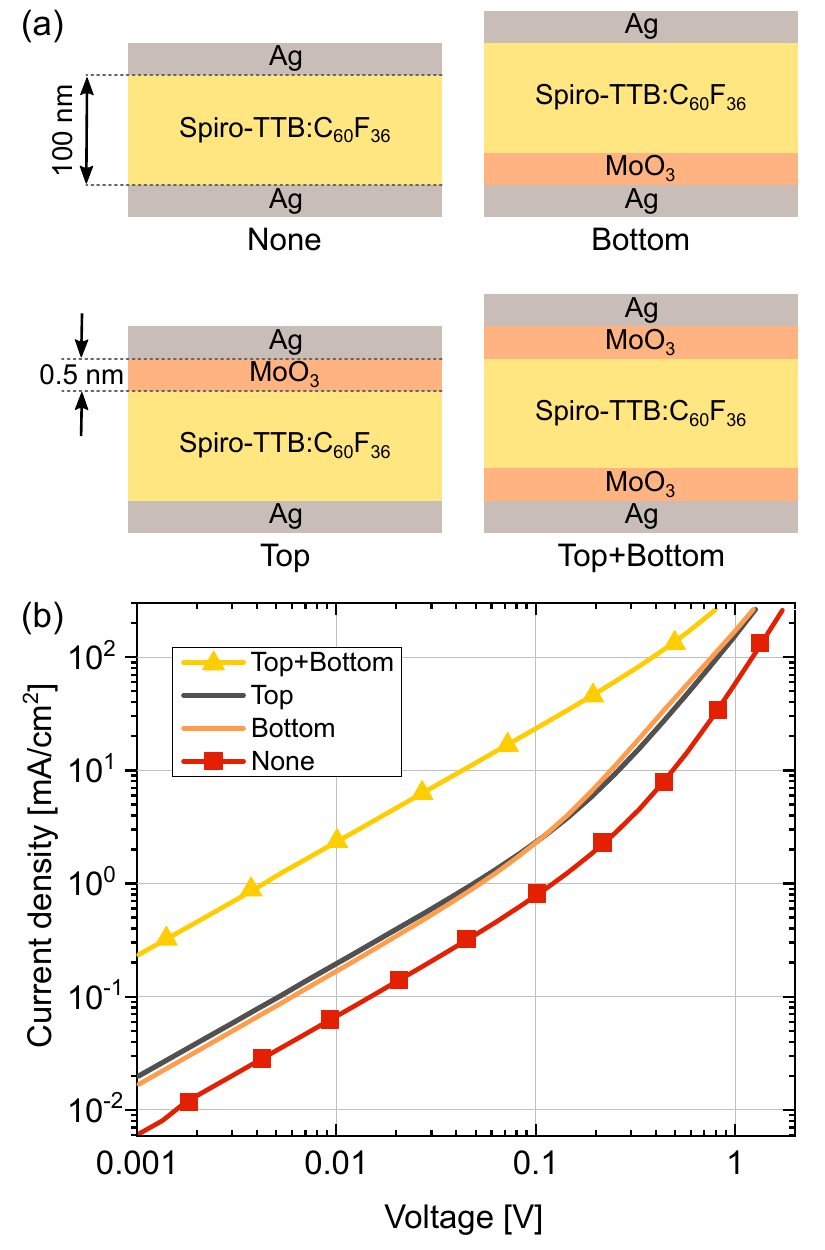}
	\caption{Panel (a) shows the investigated stack designs comprising $\SI{100}{nm}$ of p-doped organic semiconductor and either none, one, or two contact enhancement layers of \ce{MoO3} ($\SI{0.5}{nm}$). Panel (b) presents their respective $JV$ characteristics at $T=\SI{300}{K}$ and illustrates the effect on the measured contact resistance.}
	\label{figure_moox}
\end{figure}

The subsequent experiment uses an equivalently simple stack of $\SI{100}{nm}$ of Spiro-TTB doped with \ce{C60F36} ($\SI{4}{wt\percent}$) sandwiched between two silver electrodes. Again, we performed a thickness-dependent $JV$ analysis and found like for the m-MTDATA:\ce{F6-TCNNQ} devices that the resistance behavior for $J<\SI{10}{mA/cm^{2}}$ is not influenced by the organic layer thickness up to $l=\SI{200}{nm}$, cf. SI Sect. \Romannum{6}. That is, direct access to the device's contact resistance is guaranteed. Now, either at the top, the bottom, or both contact interfaces a thin layer ($\SI{0.5}{nm}$) of molybdenum oxide ($\ce{MoO3}$) is added, cf. Fig. \ref{figure_moox}(a), which is commonly used to reduce the contact resistance of p-doped organic semiconductor devices \cite{wangRoleMolybdenumOxide2008,matsushimaMarkedImprovementElectroluminescence2008,dingReducingEnergyLosses2020,meyerElectronicStructureMolybdenumoxide2011,kotadiyaUniversalStrategyOhmic2018}. The data in Fig. \ref{figure_moox}(b) proves that adding one injection interlayer clearly reduces the contact resistance of the respective interface. There appears to be no significant difference between employing the enhancement layer in the top or bottom configuration. Most likely, the non-enhanced contact governs the contact resistance characteristics here. Using \ce{MoO3} on both interfaces reduces the contact resistance at $\SI{10}{mA/cm^2}$ by more than one order of magnitude from about $\SI{30}{\Omega\,cm^2}$ to below $\SI{3}{\Omega\,cm^2}$. A detailed investigation of this material combination can be found in the SI Sect. \Romannum{6}. In such a fashion, contact resistance enhancing strategies can easily be evaluated.

\section{Conclusion}
This article presents a strategy to measure the isolated contact resistance between metals and doped organic semiconductors up to reasonable device operating conditions. For the investigated model structure of a p-doped film (\ce{m-MTDATA}:\ce{F6-TCNNQ}, $\SI{4}{wt\percent}$) between to silver electrodes, the two metal-organic interfaces cause a potential drop of about $\SI{400}{mV}$ at $\SI{10}{mA/cm^2}$. At this current density, one contact poses therefore a resistance of $R_\mathrm{contact}\approx\SI{20}{\Omega\,cm^2}$. The presented conception requires neither complex experiments nor sophisticated theoretical treatment but is based entirely on an easy-to-perform $JV$ scan. As a proof of concept, we perform a thorough evaluation for the mentioned hole-only device configuration. We employ resistance and impedance measurements while scanning a variety of device temperatures and organic film thicknesses and successfully compare the results to a simple equivalent circuit realized in LTspice. The model uses one global set of parameters replicating all experimental findings. This proves that the measured device characteristics directly disclose the contact behavior of the device up to about $\SI{100}{mA/cm^2}$ for the given system, which is sufficient for most operating scenarios in LEDs or solar cells. We further present how our measurement concept can help other groups evaluate their contact enhancement strategies by investigating the influence of molybdenum oxide as an injection layer in a second material system comprising Spiro-TTB doped with \ce{C60F36} and silver electrodes. The presented device architecture turns out to achieve extremely strong and homogeneous electric fields, which allows to specifically study the Poole-Frenkel effect in organic semiconductors.

\section{Methods}
\subsection{Device fabrication}
The hole-only devices were fabricated by thermal evaporation under high vacuum (Kurt J. Lesker Company, evaporation pressure $<\SI{1e-6}{mbar}$) on $\SI{2.5}{cm}\times \SI{2.5}{cm}$ glass substrates (Schott Borofloat33 glass, Prince Optics) of $\SI{1.1}{mm}$ thickness. The substrates were cleaned in an ultrasonic bath with acetone, ethanol, and deionized water. Aluminum (Chempur) and silver (M\&K GmbH Jena), $\SI{50}{nm}$ each, are used as bottom and top electrodes. Only the silver contacts are drawn in Fig. \ref{figure_stack_circuit} for simplicity. The p-layer consists of m-MTDATA (Synthon, sublimation cleaned) doped with \ce{F6-TCNNQ} (Novaled AG, sublimation cleaned) at a ratio of $\SI{4}{wt\percent}$, or Spiro:TTB (Lumtec, sublimation cleaned) doped with \ce{C60F36} (Ionic Liquids Technologie GmbH, sublimation cleaned) also at $\SI{4}{wt\percent}$. The thickness and deposition rates were monitored using a quartz crystal microbalance. Please find the full names of the materials and reverse $JV$ characteristics in the SI Sect. \Romannum{1} and \Romannum{3}.

To prohibit air and moisture contamination, the device stacks were encapsulated under nitrogen atmosphere after fabrication. The encapsulation glass (Sodalime glass, AMGTECH Korea) comprises a small cavity above the pixels that prevents direct contact between sensitive materials and the encapsulation glass. It was attached to the substrate using an epoxy resin (XNR5516Z-L and XNR5590, Nagase Europa GmbH).

\subsection{Device evaluation}
The devices were placed in a Peltier element-equipped cryostat that is controlled by a temperature controller (Belektronig, HAT control). The air pressure inside the cryostat was reduced to below $\SI{1}{mbar}$ using a pre-vacuum pumping system (Trivac D16B, Germany) to keep a steady temperature and preventing unwanted air convection. Every measurement was performed using the four-wire method run by a dual-channel SMU (Keithley 2602). Impedance measurements were performed using an LCR meter (Hewlett Packard 4284A precision LCR meter) and a home-built switching matrix to switch between SMU und LCR and target individual device pixels. Temperature controller, LCR meter, and SMU were run and controlled by the software tool SweepMe! \cite{fischer2016sweepme}, which enables automated measurement protocols.

\subsection{Equivalent circuit modeling}\label{sec_equiv_circuit}

The software LTspice (Linear Technology) was used for modeling the equivalent circuit presented in Fig. \ref{figure_stack_circuit}(d) with one fixed set of parameters to achieve all model results, cf. Tab. \ref{tab_paramters}. In contrast to an analytical Python code, LTspice can self-consistently iterate solutions to model voltage dependent impedance measurements, as  presented in Fig. \ref{figure_CV-thickness}.  

\begin{table}[h]
	\caption{Global parameter set used for equivalent circuit modeling with LTspice.}\label{tab_paramters}
	\begin{ruledtabular}
		\begin{tabular}{lll}
			Diode ideality factor & $n_\mathrm{i}$ & 1\\
			Tunneling scaling factor & $c$ & $\num{2e-1}$\\
			Tunneling scale conductivity & $\sigma_\mathrm{0,R}$ & $\SI{4e3}{mS\,cm^{-2}}$\\
			Diode scale current density & $j_\mathrm{0,S}$ &$\SI{5e2}{mA\,cm^{-2}}$\\
			Interface potential barrier & $V_\mathrm{BI}$ & $\SI{0.35}{V}$\\
			Scale mobility & $\mu_\mathrm{0}$ & $\SI{2e-1}{cm^2V^{-1}s^{-1}}$\\
			Scale doping concentration & $N_\mathrm{A,0}$ & $\SI{10e18}{cm^{-3}}$\\
			Relative permittivity & $\epsilon_{\mathrm{r}}$ & $\num{3.5}$\\
			Doping activation energy & $E_\mathrm{A,a}$ & $\SI{25}{meV}$\\
			Mobility activation energy & $E_\mathrm{\mu,a}$ & $\SI{0.3}{eV}$\\
		\end{tabular}
	\end{ruledtabular}
\end{table}

\bibliography{biblio_contact}

\section*{Acknowledgements}
This work was supported by the German Research Foundation (DFG) within the project HEFOS (Grant No. FI 2449/1-1) and EFOD (Grant No. RE
3198/6-1). The authors thank Andreas Wendel and Tobias Günther for manufacturing the devices.

\section{Author contributions}
A.F. conceptualized the study. A.K. and A.F. analyzed the data and performed the LTspice implementation. A.K. prepared the graphs and mainly wrote the manuscript. A.F., R.W., P.I. and K.M. set up and conducted the experiments. J.B. and S.M. contributed major ideas to the physical understanding. S.R. supervised the project. All authors discussed the results, reviewed the manuscript, and contributed to the text. 

\section{Competing interests}
Dr. Axel Fischer is co-founder of “SweepMe! GmbH” which provided the measurement software “SweepMe!” (sweep-me.net). The name of the program is given in the manuscript. The other authors declare no conflicts of interest.

\section{Materials and Correspondence}
All material, code and data is available on reasonable request.

\end{document}


\title{\Large{Supplementary information}\\\large{A simple strategy to measure the contact resistance between metals and doped organic films}}
\author{Anton Kirch}
\author{Axel Fischer}
\author{Robert Werberger}
\author{Shayan Miri Aabi Soflaa}
\affiliation{
	Dresden Integrated Center for Applied Physics and Photonic Materials (IAPP) and Institute of Applied Physics, Technische Universität Dresden, Germany\\
	Nöthnitzer Straße 61, 01187 Dresden, Germany 
}	
\author{Karolina Maleckaite}
\affiliation{
	Dresden Integrated Center for Applied Physics and Photonic Materials (IAPP) and Institute of Applied Physics, Technische Universität Dresden, Germany\\
	Nöthnitzer Straße 61, 01187 Dresden, Germany 
}	
\affiliation{Center of Physical Sciences and Technology, Sauletekio av. 3, LT-10257 Vilnius, Lithuania}
\author{Paulius Imbrasas}
\author{Johannes Benduhn}
\author{Sebastian Reineke}
\affiliation{
	Dresden Integrated Center for Applied Physics and Photonic Materials (IAPP) and Institute of Applied Physics, Technische Universität Dresden, Germany\\
	Nöthnitzer Straße 61, 01187 Dresden, Germany 
}	
\email{sebastian.reineke@tu-dresden.de}

\date{\today}
\maketitle
\clearpage

\section{Estimation of doping concentration}
The acceptor doping concentration $N_\mathrm{A}$ is a parameter fed into the equivalent circuit model. The depletion width $W$ and the transport resistance $R_\mathrm{trans}$ depend directly on $N_\mathrm{A}$, as can be extracted from the respective formulas in the main manuscript. The materials used throughout the manuscript are 4,4',4''-Tris(3-methylphenylphenylamino)triphenylamine (m-MTDATA) and the dopant 2,2'-(perfluoronaphthalene-2,6-diylidene)dimalononitrile (\ce{F6-TCNNQ}).

\begin{table}[h]
	\caption{Estimation of dopant state density}\label{table_doping_concentration}
	\begin{ruledtabular}
		\begin{tabular}{lll}
			Quantity & Value & Reference \\\hline
			m-MTDATA molar mass, $M_\mathrm{m-MTDATA}$ & $\SI{790}{g/mol}$ & \cite{MMTDATAChemSpider}\\
			m-MTDATA density, $\rho_\mathrm{m-MTDATA}$ & $\SI{1.2}{g/cm^3}$ & \cite{MMTDATAChemSpider}\\
			m-MTDATA density of states, DOS\textsubscript{m-MTDATA} & $\SI{9e20}{1/cm^3}$ &\\
			\ce{F6-TCNNQ} molar mass, $M_\mathrm{F6-TCNNQ}$ & $\SI{362}{g/mol}$ & \cite{schlesingerEnergyLevelControlHybrid2016}\\
			doping concentration, $c_\mathrm{F6-TCNNQ}$ & $\SI{4}{wt\%}$ &\\
			doping density, $\rho_\mathrm{F6-TCNNQ}$ & $\SI{0.048}{g/cm^3}$ &\\
			doping density of states, DOS\textsubscript{F6-TCNNQ} & $\SI{8e19}{1/cm^3}$ &\\
			doping efficiency $@\,\SI{4}{wt\percent}$, $\eta_\mathrm{Doping}$ & $\approx\SI{10}{\%}$ & \cite{tietzeElementaryStepsElectrical2018}\\
			Ionized acceptor density, $N_\mathrm{A}$ & $\SI{8e18}{1/cm^3}$ &
		\end{tabular}
	\end{ruledtabular}
\end{table}

Table \ref{table_doping_concentration} summarizes the values necessary to estimate $N_\mathrm{A}$. With $N_\mathrm{Av}$ being the Avogadro constant, the ionized acceptor density $N_\mathrm{A}$ can be estimated.
\begin{align}
	N_\mathrm{A}&=\eta_\mathrm{Doping}\text{DOS\textsubscript{F6-TCNNQ}}\\
	& = \eta_\mathrm{Doping}\frac{\rho_\mathrm{F6-TCNNQ}}{M_\mathrm{F6-TCNNQ}}N_\mathrm{Av}\\
	& = \eta_\mathrm{Doping}\frac{\rho_\mathrm{m-MTDATA}c_\mathrm{Doping}}{M_\mathrm{F6-TCNNQ}}N_\mathrm{Av}\\
	& \approx  \SI{8e18}{cm^{-3}}
\end{align}

As found in Ref. \cite{tietzeElementaryStepsElectrical2018}, carrier release has an activation energy of a few tens of meV. By comparing our experimental results to the equivalent circuit model, we found an activation energy of about $\SI{0.025}{eV}$ best fits our data.
\begin{align}
	N_{\mathrm{A}}=\SI{10e18}{cm^{-3}}\exp\left(\frac{\SI{-0.025}{eV}}{k_\mathrm{B}T}\right)\label{equ_doping} 
\end{align}

\begin{figure}[h]
	\includegraphics[width=0.7\linewidth]{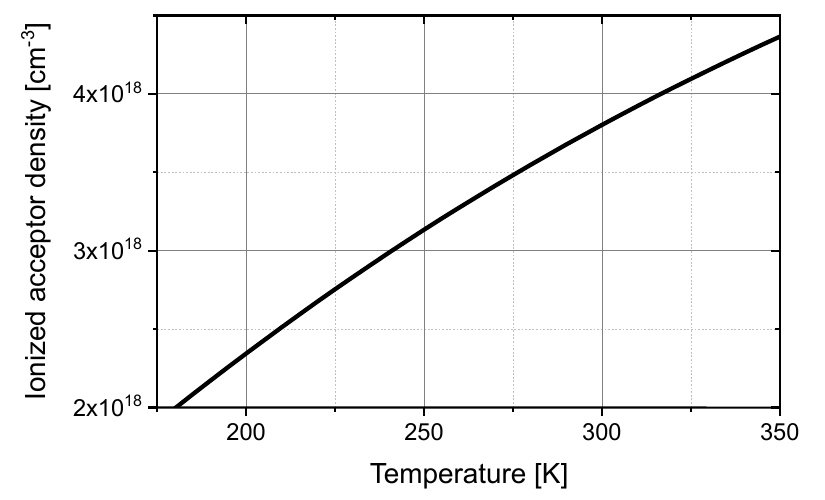}
	\caption{Temperature-dependent doping concentration according to eq. \eqref{equ_doping} as estimated for the equivalent circuit modeling.}
	\label{figure_doping_concentration}
\end{figure}

This corresponds to the temperature-dependent ionized acceptor density depicted in Fig. \ref{figure_doping_concentration}.

\clearpage
\section{Estimation of transport characteristics}
In a first approach, we considered two possible transport characteristics in the doped organic semiconductor. It could be either Ohmic or space-charge limited. To estimate the respective influences, we did the following calculation of the crossing voltage $V_\mathrm{cross}$, the voltage at which both contributions would be equal.
\begin{align}
	j_\mathrm{SCLC}&=j_\mathrm{Ohm}\\
	\frac{9}{8}\epsilon_0\epsilon_{\mathrm{r}}\mu\frac{V^2}{d^3}&=e\,N_\mathrm{A}\mu\frac{V}{d}\\
	\frac{9}{8}\epsilon_0\epsilon_{\mathrm{r}}\frac{V}{d^2}&=e\,N_\mathrm{A}\\
	V&=\frac{8}{9\epsilon_0\epsilon_{\mathrm{r}}}d^2e\,N_\mathrm{A}\label{equ_cross_voltage}\\
	V&\approx\SI{0.02}{V}\cdot d^2\quad(\text{$d$ in }[\si{nm}])\\
	V_\mathrm{cross}(d=\SI{50}{nm})&\approx\SI{50}{V}\\
	V_\mathrm{cross}(d=\SI{400}{nm})&\approx\SI{3}{kV}
\end{align}

Due to the very high doping concentration of $N_\mathrm{A}\approx\SI{4e18}{cm^{-3}}$ at room temperature, $V_\mathrm{cross}$ is well out of the investigated voltage range. It is therefore reasonable to assume Ohmic transport to be the prevailing characteristics. 

\clearpage
\section{Symmetric device behavior}
\begin{figure}[h]
	\includegraphics[width=0.7\linewidth]{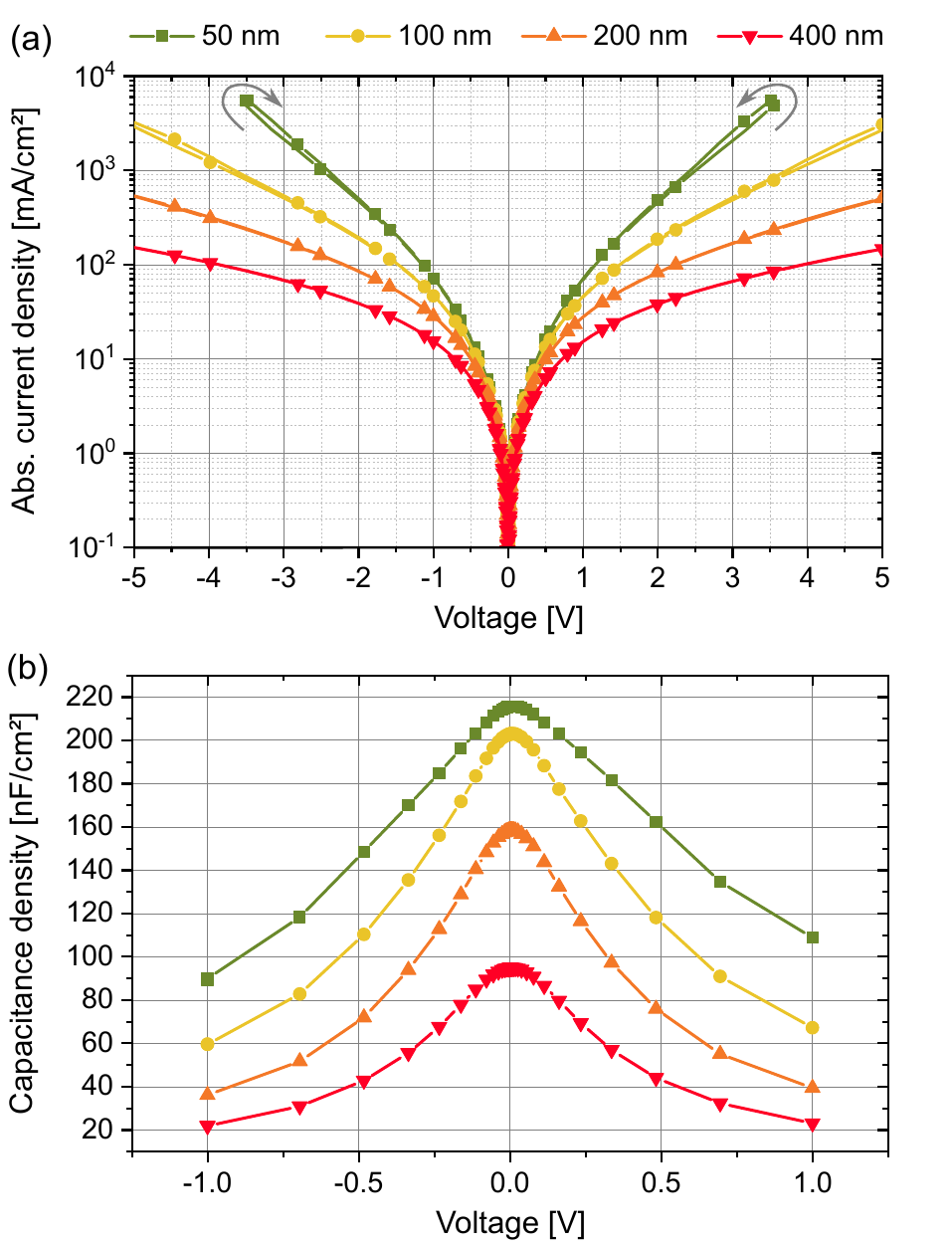}
	\caption{Panel (a) shows $JV$ and Panel (b) $CV$ characteristics for all four devices investigated in the main manuscript at $T=\SI{300}{K}$. The gray arrows in panel (a) indicate the direction of measurement.}
	\label{figure_IV_CV_symmetry}
\end{figure}

The stack design of the presented hole-only devices is symmetric. This is reflected by the $JV$ and $CV$ measurements, which are also invariant to the bias polarity, cf. Fig. \ref{figure_IV_CV_symmetry}.

The $JV$ scan is performed starting from $V=\num{0}$ to forward bias $V=\SI{5}{V}$, back to $V=\num{0}$, to the reverse bias of $V=\SI{-5}{V}$, and back to $V=\num{0}$. The $JV$ curve of the $\SI{50}{nm}$ device is limited by the current compliance of $\SI{5}{mA}$. The $CV$ measurement was run from $V=\SI{-1}{V}$ to $V=\SI{1}{V}$. Both measurements show the expected symmetric behavior.

The higher the current density of the given device, i.e. the thinner the organic film, the more pronounced is a slight hysteresis in the $JV$ scan. We account this effect to self-heating as studied e.g. in Refs. \cite{kirchExperimentalProofJoule2020,kirchElectrothermalTristabilityCauses2021}.

\clearpage
\section{Active area variation}
The produced hole-only devices show no susceptibility to their active area. We manufactured four different area sizes using four different evaporation masks and measured their respective current density. The active device area was determined using a Nikon motorized microscope Eclipse LV100ND, equipped with a NIKON high-definition DS-Fi2 camera using Nikon's \textit{NIS-Elements D} software. Pixel 4 has an active device area of more than $\SI{5}{mm^2}$. Its precise value could not be measured, as even the minimal magnification of the microscope could not display the whole active area in one picture. We, therefore, cannot give an exact $JV$ curve for that measurement and leave it out of the evaluation.

\begin{table}[h]
	\caption{Active device area}\label{table_area_variation}
		\begin{tabular}{c @\qquad c @\qquad c @\qquad c}
			\toprule\\
			Pixel 1 $[\si{mm^2}]$ & Pixel 2 $[\si{mm^2}]$ & Pixel 3 $[\si{mm^2}]$ & Pixel 4 $[\si{mm^2}]$\\
			0.09 & 0.38 & 1.23 & $> 5$\\
			\botrule\\
		\end{tabular}
\end{table}

\begin{figure}[h]
	\includegraphics[width=0.7\linewidth]{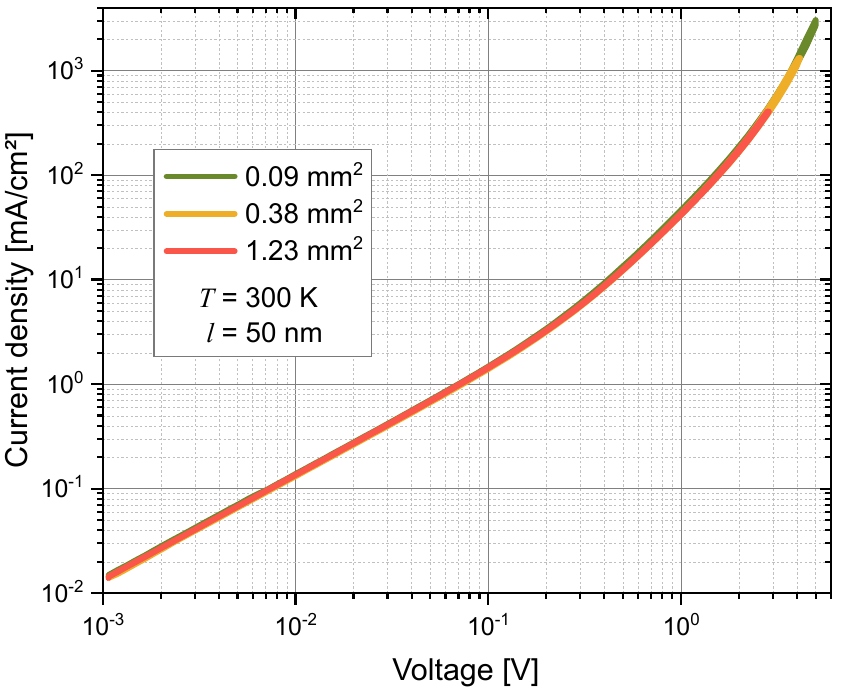}
	\caption{Variation of active device area does not show a deviation of $JV$ characteristics. The measurement was taken at $T=\SI{300}{K}$ for the $l=\SI{50}{nm}$ device.}
	\label{figure_area_variation}
\end{figure}

Figure \ref{figure_area_variation} presents the $JV$ data for Pixels 1 to 3 at $T=\SI{300}{K}$ for the $l=\SI{50}{nm}$ device. As the characteristics are perfectly overlapping and do not scale with active area size, boundary effects in the device play apparently no decisive part for our device geometry. This hints toward the devices showing homogeneous electric field distributions within the organic semiconductor.

\begin{figure}[h]
	\includegraphics[width=0.7\linewidth]{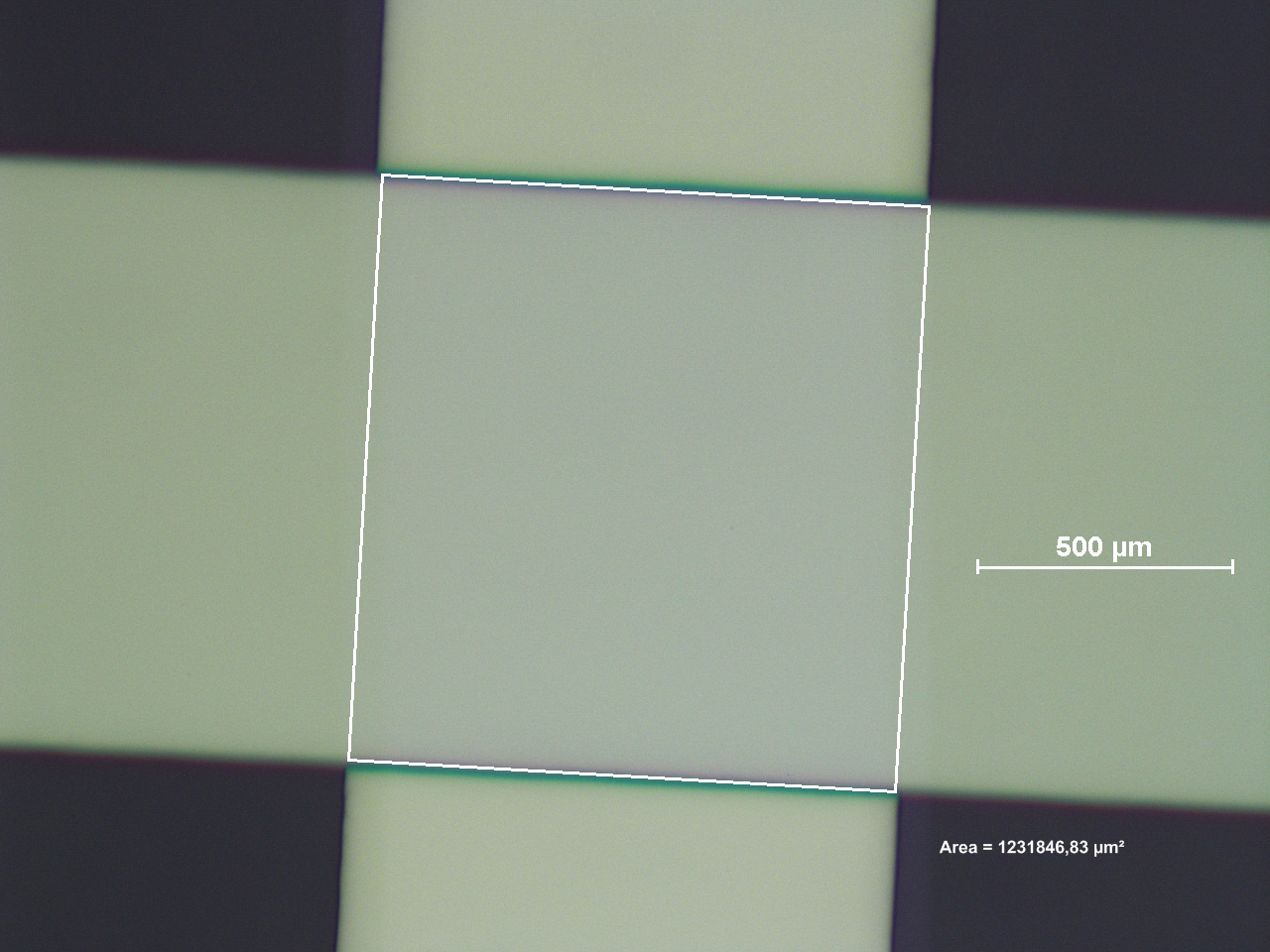}
	\caption{Microscopy image for measuring the active device area using Nikon's dimension evaluation tool \textit{NIS-Elements D}.}
	\label{figure_area_measurement}
\end{figure}

\clearpage
\section{Capacitance evaluation}
\begin{figure}[h]
	\includegraphics[width=0.4\linewidth]{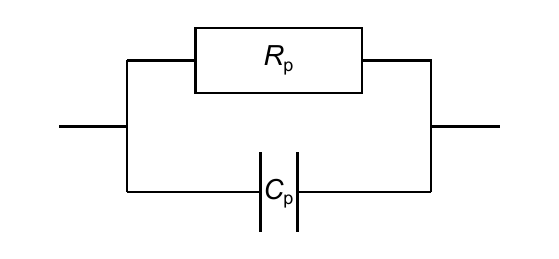}
	\caption{Parallel $RC$ element.}
	\label{figure_parallel_circuit}
\end{figure}

The four-wire measurement setup was also used to evaluate the device capacitance using an HP 4284A LCR meter. Given the frequency $\omega$ and the sinusoidal data sets of $V(t)$ and $I(t)$, which can be interpreted as complex numbers (phasors), with their respective phase shift $\varphi$, the complex impedance 
\begin{align}
	Z = |Z|e^{i\varphi} = \frac{V}{I} = \frac{|V|e^{i(\omega t+\varphi)}}{|I|e^{i\omega t}}
\end{align}
of the device can be calculated. With $X=\text{Im}(Z)$ being the reactance of the system, the system's capacitance
\begin{align}
	C_\mathrm{p} = \frac{-X}{\omega|Z|^2}
\end{align}
can be calculated. The deduction of this formula is given in the subsequent lines. Considering a parallel $RC$ circuit as given in Fig. \ref{figure_parallel_circuit}, the impedance $Z$ can be calculated with $X_\mathrm{p}$ being the reactance of the parallel capacitor with capacity $C_\mathrm{p}$.

\begin{align}
	Z & = \left(\frac{1}{X_\mathrm{p}}+\frac{1}{R_\mathrm{p}}\right)^{-1}=\frac{R_\mathrm{p}X_\mathrm{p}}{R_\mathrm{p}+X_\mathrm{p}},\qquad X_\mathrm{p}=\frac{-i}{\omega C_\mathrm{p}}\\
	& = \frac{\frac{-iR_\mathrm{p}}{\omega C_\mathrm{p}}}{R-\frac{i}{\omega C_\mathrm{p}}}=\frac{\frac{R_\mathrm{p}}{\omega^2C_\mathrm{p}^2}-\frac{iR_\mathrm{p}^2}{\omega C_\mathrm{p}}}{R_\mathrm{p}^2+\frac{1}{\omega^2 C_\mathrm{p}^2}}\\
	X & = \text{Im}(Z) = \frac{-\frac{R_\mathrm{p}^2}{\omega C_\mathrm{p}}}{\frac{R_\mathrm{p}^2}{\omega^2 C_\mathrm{p}^2}\left(\omega^2C_\mathrm{p}^2+\frac{1}{R_\mathrm{p}^2}\right)} = \frac{-\frac{R_\mathrm{p}^2}{\omega C_\mathrm{p}}}{\frac{R_\mathrm{p}^2}{\omega^2 C_\mathrm{p}^2}\frac{1}{|Z|^2}}=-\omega C_\mathrm{p} |Z|^2\\
	C_\mathrm{p} & = \frac{-X}{\omega |Z|^2} = \frac{-\text{Im}\left(\frac{V}{I}\right)}{\omega \left|\left(\frac{V}{I}\right)\right|^2}  
\end{align}

So, given that $V(t)$ and $I(t)$ are interpreted as complex numbers, the device capacity can be evaluated as a function of the experimentally accessible quantities $V(t)$, $I(t)$, and $\omega$.

\clearpage
\section{Investigation of molybdenum oxide as contact interlayer}

In the main manuscript, the experimental conception of measuring the contact resistance of a given metal/doped organic film interface is briefly presented for a series employing molybdenum oxide (\ce{MoO3}) as additional interlayer. 

\begin{figure}[h]
	\includegraphics[width=0.9\linewidth]{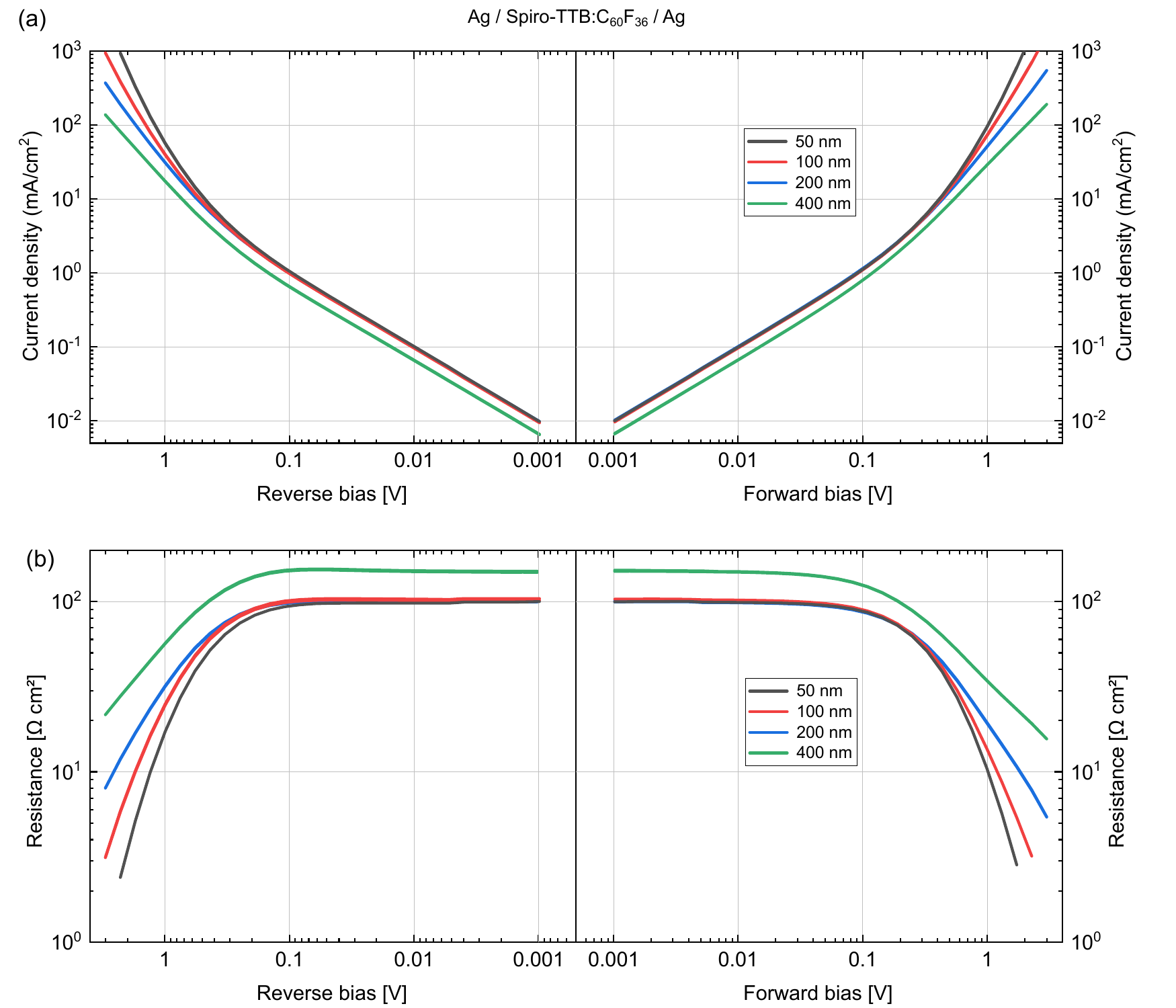}
	\caption{Panel (a) shows the $JV$ and (b) the $RV$ characteristics of the Ag/Spiro-TTB:\ce{C60F36}/Ag devices.}
	\label{figure_IV_Spiro}
\end{figure}

Figure \ref{figure_IV_Spiro} presents the experimental data obtained for the stack containing 2,2',7,7'-tetra(N,N-di-tolyl)amino-9,9-spirobifluorene (Spiro-TTB) doped with \ce{C60F36} ($\SI{4}{wt\percent}$) sandwiched between silver electrodes, as introduced in the main manuscript. As is also measured for the material combination in the main manuscript, the device resistance is merely influenced by a transport layer thickness of more than $l=\SI{200}{\nano m}$ up to a current density of $J=\SI{10}{mA/cm^2}$. Above that current density, the $\SI{50}{nm}$ is expected to longest follow the pure contact characteristics. 

\clearpage

\begin{figure}[h]
	\includegraphics[width=0.9\linewidth]{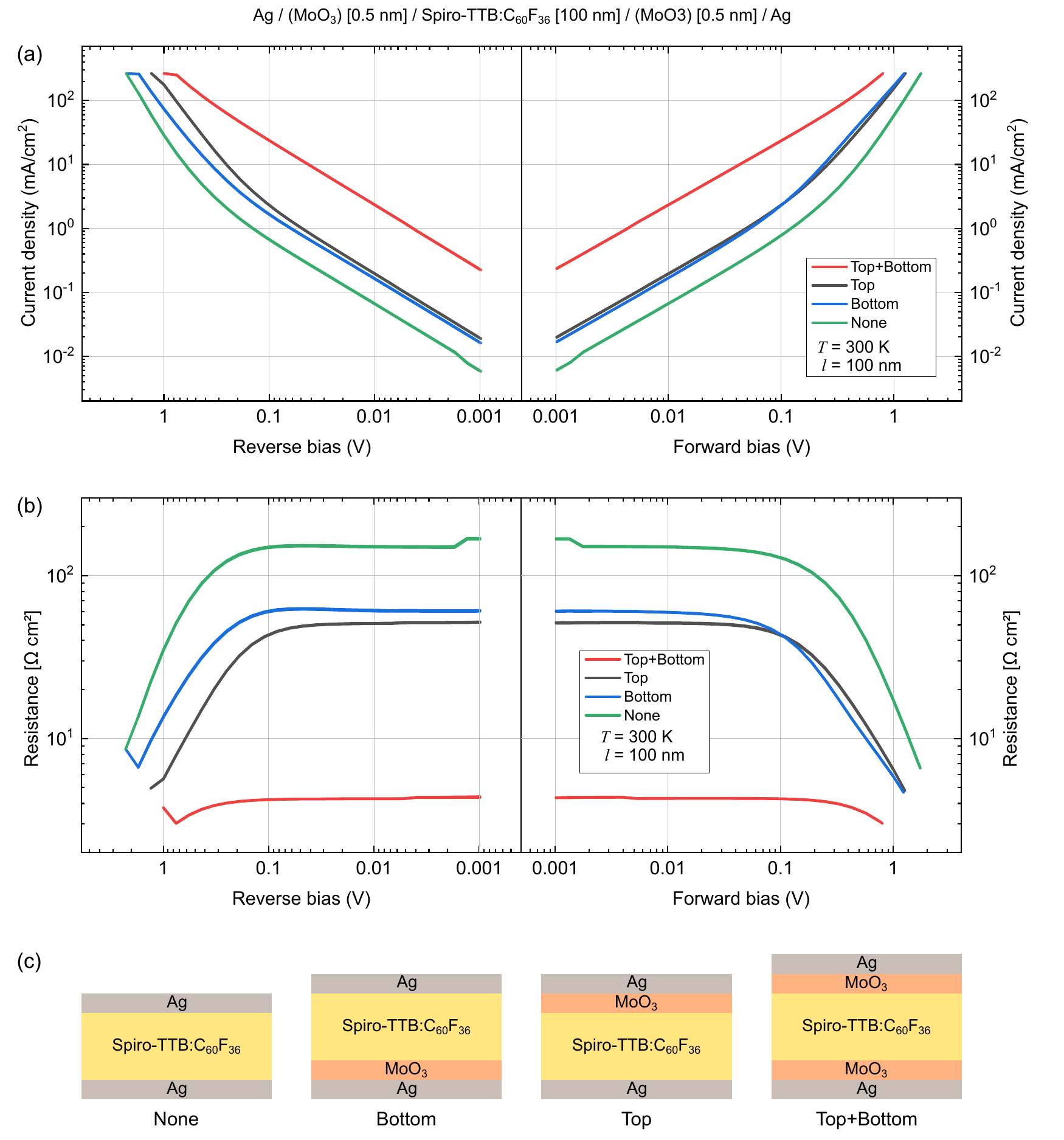}
	\caption{Panel (a) shows the $JV$ and (b) the $RV$ characteristics of a Ag/(\ce{MoO3})[$\SI{0.5}{nm}$]/Spiro-TTB:\ce{C60F36}[$\SI{100}{nm}$]/(\ce{MoO3})[$\SI{0.5}{nm}$]/Ag device. The device structures are schematically depicted in (c).}
	\label{figure_IV_Spiro_Moox}
\end{figure}

Figures \ref{figure_IV_Spiro_Moox}(a) and \ref{figure_IV_Spiro_Moox}(b) present the $JV$ characteristics of the devices displayed in Fig. \ref{figure_IV_Spiro_Moox}(c). Either at none, one (top or bottom), or both contact interfaces a thin layer ($\SI{0.5}{nm}$) of $\ce{MoO3}$ is added, which is commonly used to reduce the contact resistance of p-doped organic semiconductor devices \cite{wangRoleMolybdenumOxide2008,matsushimaMarkedImprovementElectroluminescence2008}. Figures \ref{figure_IV_Spiro_Moox}(a) and (b) demonstrates that adding such an injection interlayer clearly reduces the contact resistance. In such a fashion, contact resistance enhancing strategies can easily be evaluated. 

\begin{figure}[h]
	\includegraphics[width=0.9\linewidth]{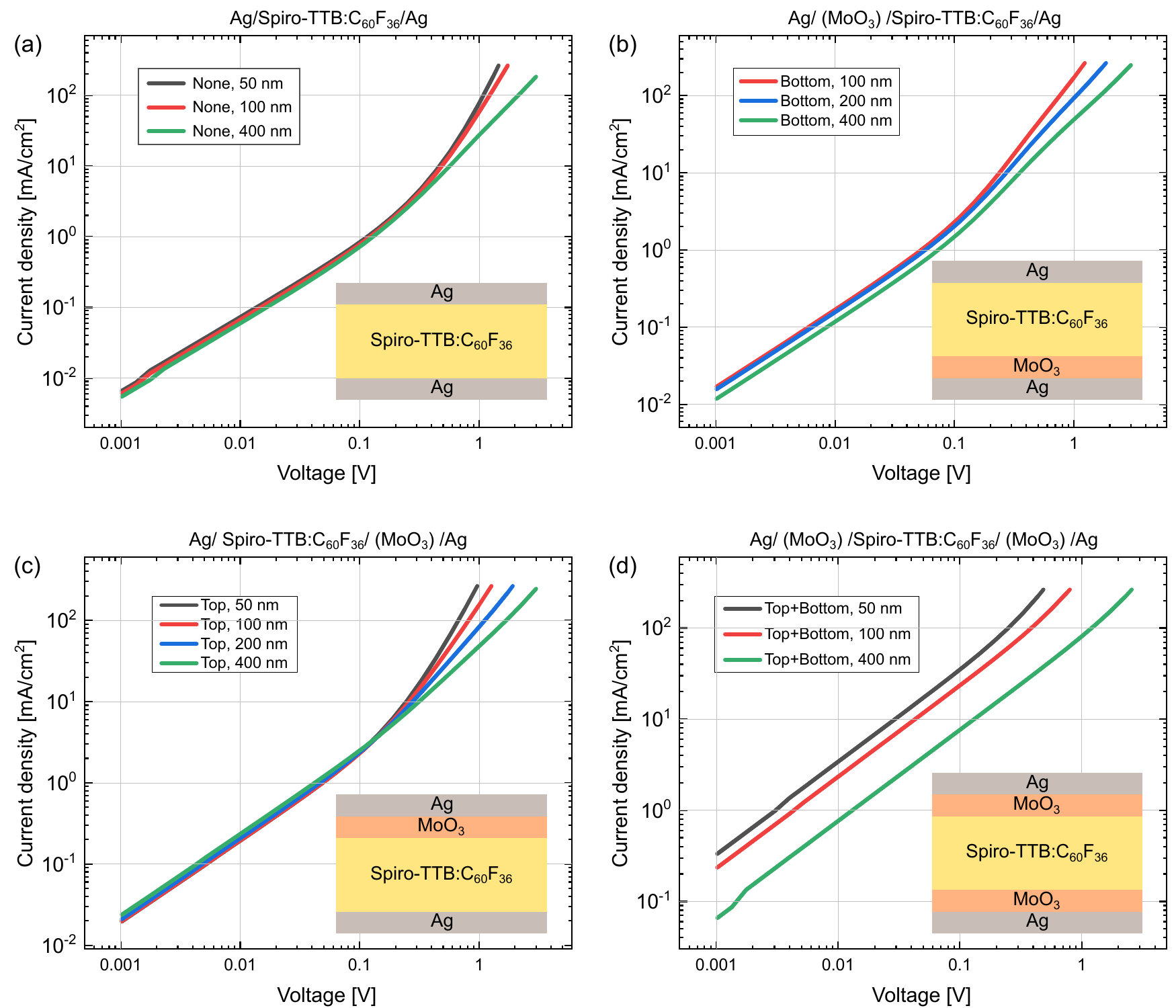}
	\caption{The $JV$ (a) and $RV$ (b) characteristics of a Ag/(\ce{MoO3})/Spiro-TTB:\ce{C60F36}/(\ce{MoO3})/Ag device. }
	\label{figure_IV_Spiro_Thickness}
\end{figure}

\clearpage
From Fig. \ref{figure_IV_Spiro_Moox} it is not directly clear whether all $JV$ curves still access the pure contact resistance. Therefore, Fig. \ref{figure_IV_Spiro_Thickness} presents the thickness-dependent $JV$ curves for all four scenarios. For none and one $\ce{MoO3}$ layer, all device characteristics overlay until they separate according to their organic film thickness at high voltages (transport characteristics take over). For the device with two $\ce{MoO3}$ layers, however, this is not the case. The $JV$ characteristics is even at low voltages already influenced by the organic's transport resistance, as the lines are always separated according to the layer thickness due to the contact resistance being even lower. The respective red ``Top+Bottom'' line in Fig. \ref{figure_IV_Spiro_Moox}(b) therefore can only be interpreted as upper boundary of the contact resistance. Its true value is even smaller. 

Figure \ref{figure_IV_Spiro_Thickness} does not sport all $4\times 4$ $JV$ characteristics, as the missing devices had short-cuts. 

\clearpage
\section{Model resistance}
\begin{figure}[h]
	\includegraphics[width=0.6\linewidth]{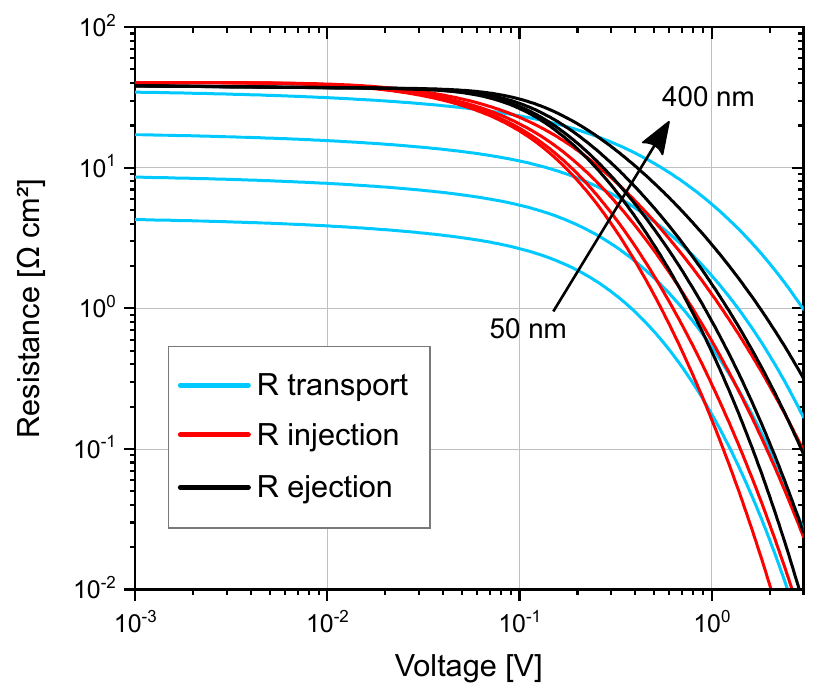}
	\caption{Modeled resistance dependencies of $R_\mathrm{inj}$, $R_\mathrm{ejec}$, and $R_\mathrm{trans}$ for different organic film thicknesses $l=\SI{50}{nm},\SI{100}{nm},\SI{200}{nm}$, and $\SI{400}{nm}$.}
	\label{figure_RV_thickness_model}
\end{figure}

From the equivalent circuit modeling a detailed analysis of the resistance contributions can be performed. Figure \ref{figure_RV_thickness_model} presents the modeled resistance contributions $R_\mathrm{inj}$, $R_\mathrm{ejec}$, and $R_\mathrm{trans}$ for all for organic film thicknesses. The color coding only separates the respective resistance types to keep the graph clear. The thickness separation is such that the thinnest layer causes both the lowest transport and contact resistance. The main message of the Figure is, first, that the contact resistance contributions dominate at low voltages. At increasing voltage, their exponential resistance behavior makes them drop below the transport resistance. This model result justifies the common assumption of contacts being ``Ohmic'', i.e. not contributing a significant series resistance, at standard operating parameters. It is plain, however, that the pure contact resistance can be accessed at low voltages. Second, the injection and ejection characteristics look very much alike, owing to their mutual exponential nature. It is therefore reasonable to approximate the injection or ejection resistance to be about half the contact resistance at low to medium voltages.

The drop in $R_\mathrm{trans}$ with increasing voltage is induced by the Poole-Frenkel (P-F) effect. A model not including P-F, would yield constant $R_\mathrm{trans}$ lines for every layer thickness, which does not follow the experimental observation.

\clearpage

\bibliography{biblio_contact}